\documentclass[aps, prl, reprint,amsmath,amssymb]{revtex4-2}

\usepackage[]{hyperref}
\hypersetup{
    colorlinks=true,
    linkcolor=blue,
    filecolor=magenta,      
    urlcolor=cyan,
    citecolor = blue
    }

\usepackage{graphicx}
\usepackage{dcolumn}
\usepackage{bm}
\usepackage{xcolor}
\usepackage{comment}

\begin{document}


\title{Nanomechanical sensor resolving impulsive forces below its zero-point fluctuations}


\author{Martynas Skrabulis$^{1,2}$}
\author{Martin Colombano Sosa$^{1,2}$}
\author{Nicola Carlon Zambon$^{3}$}
\author{Andrei Militaru$^{4}$}
\author{Massimiliano Rossi$^{5}$}
\author{Martin Frimmer$^{1,2}$}
\author{Lukas Novotny$^{1,2}$}

\affiliation{$^{1}$Photonics Laboratory, ETH Zürich, 8093 Zürich, Switzerland}
\affiliation{$^{2}$Quantum Center, ETH Zürich, 8093 Zürich, Switzerland}
\affiliation{$^{3}$Università di Padova, Dipartimento di Fisica e Astronomia, 35134 Padova, Italy}
\affiliation{$^{4}$Institute of Science and Technology
Austria, Am Campus 1, 3400 Klosterneuburg, Austria}
\affiliation{$^{5}$Kavli Institute of Nanoscience, Department of Quantum Nanoscience,
TU Delft, 2628CJ Delft, The Netherlands}

\begin{abstract}

The sensitivity of a mechanical transducer is ultimately limited by its inherent quantum fluctuations.
Here, we use an optically levitated nanoparticle to measure impulsive forces smaller than the particle's zero-point momentum uncertainty. 
Our approach relies on reversibly squeezing the levitated particle's center-of-mass motion to coherently amplify the perturbation. We demonstrate resolving single impulsive-force kicks as small as 6.9 keV/c, a value 0.6 dB below the sensor's zero-point value.

\end{abstract}

\maketitle

\paragraph{Introduction.}
Mechanical transducers have played a pivotal role in unraveling the fundamental laws of nature~\cite{Braginsky_MeasurementWeak}. 
Hallmark achievements based on mechanical transduction span from imaging individual molecular bonds in atomic force microscopy~\cite{OteyzaScience_2013_direct} to the detection of gravitational waves in kilometer-scale interferometers~\cite{Abbot_PRL_2016_gravitationalwaves}. 

The key challenge associated with mechanical transduction is to measure the motion generated by the perturbation~\cite{Braginskii_Book_QuantumMeasurement}.
Optical readout schemes encode the transducer position into the phase of an electromagnetic probe field. 
The field's quantum-mechanical nature limits the minimum readout noise via two contributions: measurement backaction and measurement imprecision~\cite{Giovannetti2004_ScienceQuantumEnhanced}. 
Optimally balancing these two contributions leads to the ``standard quantum limit'' of continuous displacement and force sensing~\cite{CavesRevModPHys_1980_measurement,ClerkRevMovPhys_2010,Clerk_PRB_2004_quantumLimited,aspelmeyer_RMP_2014_optomechanicsreview}. 
At frequencies different from the oscillator's resonance, this standard limit can be overcome by exploiting quantum correlations between the amplitude and phase quadrature of the probe after it interacted with the transducer (termed variational readout~\cite{Clerk_PRB_2004_quantumLimited,Safavi-Naeini2013_squeezed,Purdy2013_PRXstrong,Mason_natphys_2019_belowsql,kampel_PRX_2017_variationalreadout}), or by using squeezed probe light with engineered quantum correlations~\cite{Caves1981_PRD_quantumNoiseInterf,Jia2024_Science_squeezing}.
The resulting (ultimate) quantum limit is given by the inevitable zero-point motion of the transducer and poses a hard constraint for continuous position and force sensing~\cite{Clerk_PRB_2004_quantumLimited,Giovannetti2004_ScienceQuantumEnhanced}.

In recent years, instead of focusing on the probe field, engineered quantum correlations of the transducer's motion have been leveraged~\cite{McCormick2019_Nature_quantumEnh,gilmore_science_2021_quantumenhanced,colombo2022time} for sensing impulsive forces, i.e., instantaneous perturbations of the transducer's momentum.  
Noiseless amplification by squeezing the vacuum fluctuations of the transducer has been demonstrated with ions~\cite{Burd_Science_2019_CMA}, which resemble microscopic mechanical oscillators when confined in radio-frequency traps.

It is attractive to transfer such coherent amplification protocols~\cite{Caves1981_PRD_quantumNoiseInterf,Hudelist_natcomm_2014_quantummetrology,Burd_Science_2019_CMA} to much more massive systems. A prime target is an optically trapped dielectric nanoparticle ($\sim10^8$ atomic mass units)~\cite{Novotny_science_2021_levitodynamics_review}, which is a promising candidate to contribute to refinements of the standard model~\cite{Moore_2025_PRXQ_DMsearch,moore_QST_2021_newphysics,Carney_PRXQ_2023_neutrinosearch}. 
Proposals envision detecting collisions with hypothetical dark-matter particles~\cite{Moore_PRL_2020_compDM, carney_QST_2011_mechanicalsensing}, or unknown products of nuclear decays, with experimental efforts already underway~\cite{wang_PRL_2024_nucleardecays}.

Quantum control of levitated systems, a field termed levitodynamics, has made tremendous progress in recent years~\cite{Novotny_science_2021_levitodynamics_review}.
Optical trapping in ultra-high vacuum and efficient quantum measurement of the nanoparticle's center-of-mass motion have enabled ground-state cooling using active feedback~\cite{magrini_nature_2021_quantumcontrol,tebbenjohanns_nature_2021_groundstate,Kamba2022optical} and cavity cooling~\cite{delic_science_2020_cavitygs,piotrowski2023simultaneous,Ranfagni2022twodimensional}. Furthermore, rapid modulation of the optical trapping potential has enabled the generation of squeezed states of mechanical motion~\cite{Kamba_Science_2025_Quantumsqueezing,Rossi_PRL_2025_QuantumDeloc}. 
In this situation, it is enticing to exploit these quantum correlations to boost sensitivity to impulsive forces.

In this work, we resolve impulsive forces acting on an optically levitated nanoparticle which are smaller than the particle's zero-point fluctuations.
We realize a coherent amplification protocol that relies on reversible and coherent squeezing operations, which we implement by rapidly switching the stiffness of the optical potential trapping the particle. 
Our system can resolve impulsive forces as small as $6.9\pm0.8$~keV/c, which is $0.6^{+0.6}_{-0.4}$~dB below the particle's zero-point momentum fluctuations.

\paragraph{Key concept.}

\begin{figure}
\includegraphics[width=\columnwidth]{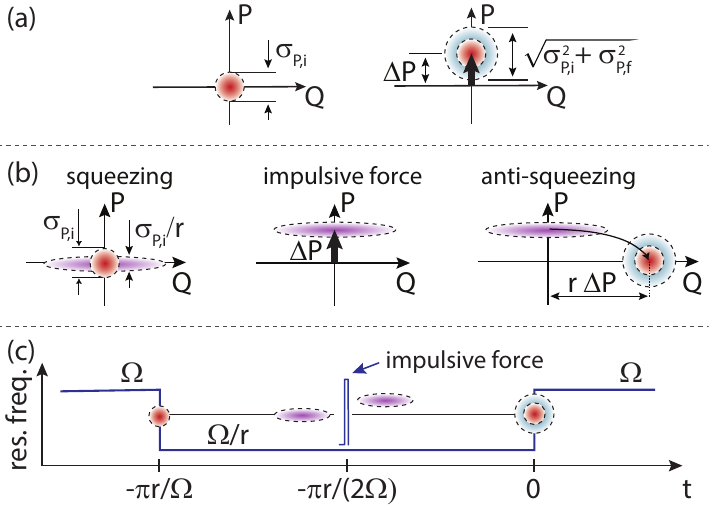}
\caption{\label{fig:concept} 
(a)~Conventional impulsive-force sensing. 
Left: A harmonic oscillator is initialized to a Gaussian state centered in phase-space with momentum uncertainty $\sigma_{P,i}$ (red).
Right: An impulsive force translates the state by $\Delta P$ along the momentum axis. Reading out the displaced state adds estimation uncertainty $\sigma_{P,f}$ (blue).
(b)~Coherent amplification. 
Left: The harmonic oscillator is initialized as in (a). The momentum uncertainty is squeezed to $\sigma_{P,i}/r$ (purple ellipse). 
Center: The impulsive force displaces the state by $\Delta P$ along the momentum axis.
Right: Anti-squeezing turns the momentum displacement $\Delta P$ into a position displacement $\Delta Q=r\Delta P$. 
(c)~Modulation sequence for oscillator resonance frequency. The frequency is temporarily stepped from $\Omega$ to the reduced value $\Omega/r$ to squeeze and subsequently anti-squeeze the oscillator. The impulsive force acts at the time of maximum momentum squeezing.
}
\end{figure}

Let us revisit the standard protocol for impulsive force sensing. 
A harmonic oscillator (mass $m$, eigenfrequency $\Omega$) with position $z$ and momentum $p$ is initialized in a Gaussian state, centered at the origin of phase space, as illustrated in Fig.~\ref{fig:concept}(a). Let $Q=z/z_\text{zp}$ and $P=p/p_\text{zp}$ be the phase-space coordinates normalized by their respective zero-point values $z_\text{zp}=\sqrt{\hbar/(2m\Omega)}$ and $p_\text{zp}=\sqrt{\hbar m\Omega/2}$. Let the initial position variance $\sigma_{Q,i}^2$ and momentum variance $\sigma_{P,i}^2$ be equal [red in Fig.~\ref{fig:concept}(a)].
An impulsive force acting on the oscillator shifts the state along the momentum axis by the transferred momentum $\Delta P$. 
The smallest measurable $\Delta P$ is given by the total variance $\sigma^2_{P,\text{tot}}=\sigma_{P,i}^2 + \sigma_{P,f}^2$, where the second term represents the uncertainty with which the final state can be localized in phase space by subsequent measurements, illustrated in blue in Fig.~\ref{fig:concept}(a). 

We now turn to describing the coherent amplification protocol~\cite{Caves1981_PRD_quantumNoiseInterf,Hudelist_natcomm_2014_quantummetrology,Burd_Science_2019_CMA} adapted to a levitated oscillator system. 
We modulate oscillator's eigenfrequency in a step-like fashion as illustrated in Fig.~\ref{fig:concept}(c). With the oscillator initialized as in the standard protocol, we step the eigenfrequency from the initial value $\Omega$ to $\Omega/r$, with $r>1$, at time $t=-\pi r/\Omega$. In the softer potential, points in phase space move on ellipses elongated along the position axis. Specifically, at time $t=-\pi r/(2\Omega)$, the initially circular state is maximally squeezed along the momentum axis, with variance $\sigma_{P,i}^2/r^2$, and anti-squeezed along the position axis, with variance $r^2\,\sigma_{Q,i}^2$, as illustrated in Fig.~\ref{fig:concept}(b, left).
An impulsive force arriving at $t=-\pi r/(2\Omega)$ displaces this squeezed state along the momentum axis by $\Delta P$ (center panel). 
Further evolution in the weaker potential for another quarter period, until $t=0$, reverses the squeezing operation, leaving the state with the initial variances  $\sigma_{P,i}^2=\sigma_{Q,i}^2$ (right). 
Importantly, the momentum displacement $\Delta P$ is turned into a position displacement $\Delta Q = r\, \Delta P$, which is boosted by $r$ during the anti-squeezing step. 
Therefore, embedding the impulsive force between the squeezing and anti-squeezing operations amplifies the signal produced by the impulsive force without adding additional uncertainty.
In the following, we experimentally realize this coherent amplification protocol with an optically levitated nanoparticle.

\paragraph{Experimental setup.}

\begin{figure}[b]
\includegraphics[width=65ex]{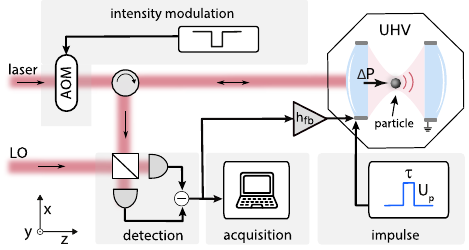}
\caption{\label{fig:setup} 
Experimental setup. A focused laser generates an optical trap for a dielectric nanoparticle in ultra-high vacuum (UHV). The backscattered light is mixed with a local oscillator (LO) to detect the particle motion along the $z$ axis. This position signal is used for feedback (FB) cooling. The optical trapping potential is modulated with an acousto-optic modulator (AOM). An impulsive force $\Delta P$ is imparted by applying a voltage pulse of duration $\tau$ and amplitude $U_p$ to metallic electrodes surrounding the nanoparticle.}
\end{figure}

\begin{figure*}[hbt]
\includegraphics[width=0.95\textwidth]{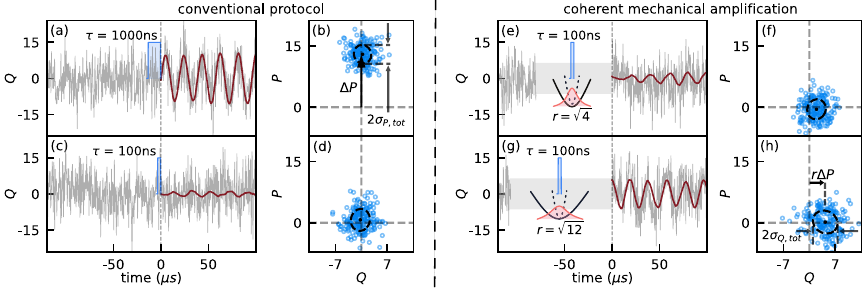}
\caption{\label{fig:timetraces} 
Conventional protocol (a--d) vs. coherent mechanical amplification (e--h). 
(a)~Gray: Unfiltered time trace of measured position $Q(t)$. 
An impulsive force (duration $\tau=1000$~ns, illustrated in blue, not to scale) is applied at $t=0$. 
Red: Filtered timetrace. 
(b)~State-space distribution right after application of the impulsive force for 200 realizations of the protocol as shown in~(a). 
Dashed ellipse indicates the covariance matrix of the distribution. 
The displacement of the state by $\Delta P$ is visible.
(c,\,d)~Same as (a,\,b), but with $\tau=100$~ns. The displacement of the state is not discernible against the uncertainty.
(e)~Time trace of a coherent amplification protocol with squeezing factor $r=\sqrt{4}$. The impulsive force ($\tau=100~$ns) has the same strength as in (c,\,d). 
(f)~State-space distribution obtained from 200 repetitions of the protocol shown in~(e). A slight displacement of the state along the position axis is discernible.
(g,\,h)~Same as (e,\,f), but with squeezing factor $r=\sqrt{12}$. The position displacement in phase-space is visibly amplified by the squeezing factor as compared to (d).}
\end{figure*}

Our impulsive-force detector is an optically levitated silica nanoparticle (nominal diameter 100 nm, mass 1.2 fg). A simplified experimental setup is shown in Fig.~\ref{fig:setup} (see Supplemental Material~\cite{supplement}).
The optical trap is generated by focusing a laser (wavelength 1550~nm, power $\sim$ 600~mW) with an aspheric lens (numerical aperture~0.8). 
The laser is linearly polarized along $x$ and propagates along $z$. 
For small displacements, the center-of-mass motion of the particle is harmonic along the three axes, with characteristic frequencies $(\Omega_z,\Omega_x,\Omega_y)/(2 \pi) =(52, 141, 175$) kHz. 
In this work, we focus on the particle motion along $z$ and drop the subscript ($\Omega=\Omega_z$) for brevity. 
We operate the optical tweezer in cryogenic ultra-high vacuum (UHV), where heating of the particle motion due to collisions with background gas molecules is negligible~\cite{tebbenjohanns_nature_2021_groundstate}.
The particle's motion imprints a position-dependent phase on the scattered laser light~\cite{tebbenjohanns_PRA_2019_optimaldetection}, which can be detected interferometrically with a detection efficiency $\eta=0.14 \pm0.02$. 
Measurement backaction leads to a phonon heating rate $\Gamma_\text{qb}/(2\pi)= \text{3.4}\pm \text{0.5}    \,\text{kHz}$.
We use the position measurement to implement a cold-damping feedback~\cite{tebbenjohanns_PRL_2019_colddamping} that reduces the phonon occupation of the particle's $z$ mode to $n = 1.2 \pm 0.6$ quanta (see Supplemental Material~\cite{supplement}).
The feedback force is generated by applying a voltage to the metallic lens holders of the optical trap. The resulting electric field acts on the charge carried by the particle (typical charge-to-mass ratio 1~C/kg)~\cite{Frimmer2017_netCharge}.

We furthermore harness the particle's charge to apply controlled impulsive forces. To this end, we send voltage pulses of constant amplitude $U_p=2~$V to the electrodes, as illustrated in Fig.~\ref{fig:setup}. 
We tune the transferred momentum via the pulse duration $\tau$ which is restricted to $\tau\le 1~\mu$s in all experiments, much shorter than the particle's oscillation period ($\approx 20~\mu$s). Therefore, the momentum transferred to the nanoparticle is linear in $\tau$.

\paragraph{Conventional impulsive-force sensing.}
To illustrate the effect of an impulsive force on the levitated particle, Fig.~\ref{fig:timetraces}(a) shows in gray the raw measured particle position $Q(t)$ as a function of time $t$. 
The particle is initialized by feedback cooling, which is released at $t=-3.6~\mu$s.
A pulse of duration $\tau=1000\,$ns is applied at time $t=0$ (pulse illustrated in blue, not to scale). 
The momentum $\Delta P$ transferred to the particle by the pulse leads to a visibly increased oscillation amplitude at times $t>0$.
To suppress the noise associated with the full detector bandwidth, we estimate the state of the particle after the impulsive force ($t>0$) using the optimal retrodiction filter~\cite{Rossi_PRL_2019_quantumtrajectory}. The filtered timetrace is shown in red in Fig.~\ref{fig:timetraces}(a) and provides the best estimate of the position $Q$ and momentum $P$ of the particle right after the application of the impulsive force.

To learn about the phase-space distribution of the oscillator after application of the impulsive force, we repeat the protocol 200 times. For each repetition, we plot the estimated state of the oscillator as a point in phase space in Fig.~\ref{fig:timetraces}(b). The displacement of the mean of the state by the impulsive force $\Delta P$ along the momentum axis is clearly visible. The momentum spread of the datapoints resembles the total uncertainty $\sigma_{P,\text{tot}}$.

In Figs.~\ref{fig:timetraces}(c,\,d), we repeat the measurements shown in (a,\,b), albeit with an impulsive force reduced by a factor ten ($\tau=100~$ns). The observed phase-space displacement of the state due to the impulsive force is smaller than the spread of the datapoints, i.e, the uncertainty $\sigma_{P,\text{tot}}$ with which the oscillator can be localized in phase space in a single repetition of the experiment. 

\paragraph{Coherent mechanical amplification.}
In Fig.~\ref{fig:timetraces}(e), we show the timetrace $Q(t)$ of the particle under a coherent mechanical amplification protocol with a squeezing factor $r=\sqrt{4}$ and an impulsive force quantified by $\tau=100~$ns [same as in (c,\,d)]. 
We set $t=0$ to coincide with the end of the protocol, where anti-squeezing is completed and the oscillator frequency is restored to the original value $\Omega$. During the squeezing and anti-squeezing steps, no position record is available due to the laser intensity modulation according to the sequence shown in Fig.~\ref{fig:concept}(c). To mitigate the impact of stray-field fluctuations, we deploy a decoupling protocol similar to the one described in Ref.~\cite{Rossi_PRL_2025_QuantumDeloc}.

As in the conventional protocol, we filter the data [red line in Fig.~\ref{fig:timetraces}(e)] and build the phase-space distribution shown in Fig.~\ref{fig:timetraces}(f) from 200 repetitions. While a slight displacement of the state along the position axis is visible, as expected for the amplification protocol [c.f.\ Fig.~\ref{fig:concept}(b)], the total position noise $\sigma_{Q,\text{tot}}$ is still larger than this displacement. 
To further amplify the signal, we increase the squeezing factor to $r=\sqrt{12}$ in Figs.~\ref{fig:timetraces}(g,\,h). Now, the phase-space displacement along the position axis is larger than the uncertainty $\sigma_{Q,\text{tot}}$. 
The direct comparison between Figs.~\ref{fig:timetraces}(d) and (h) shows the power of coherent mechanical amplification, where the displacement of the state is amplified by the reversible squeezing operation in an (ideally) noiseless fashion.

\paragraph{Scaling of amplification.} 
\begin{figure}[!t]
\hspace{-2ex}
\includegraphics[width=\columnwidth]{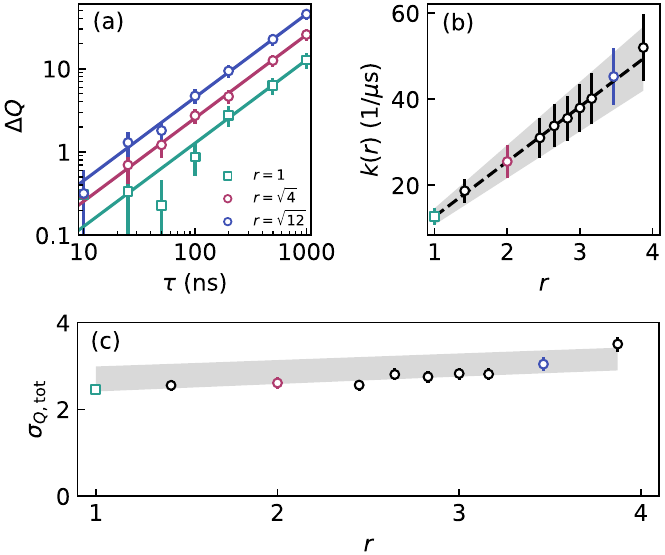}
\caption{\label{fig:eval} 
(a)~Phase-space displacement $\Delta Q$ as a function of pulse duration $\tau$ for three different squeezing ratios $r$. Solid lines: Linear fits $\Delta Q =  k(r)\,\tau$. 
Error bars: $1\sigma$ uncertainty obtained via error propagation. 
(b)~Slope $k(r)$ as a function of squeezing ratio $r$ extracted from data as shown in (a). 
Dashed line indicates $k(r) = k_1\,r$.
Shaded area: $1\sigma$ confidence interval of $k_1$. 
(c)~Total measured position standard deviation $\sigma_{Q,\text{tot}}$ as a function of $r$.
The slight increase with $r$ is due to measurement backaction.
Colored datapoints in (b) and (c) correspond to datasets shown in (a). Shaded area: $1\sigma$ confidence interval of the model~\cite{supplement}.
}
\end{figure}
To further corroborate our understanding of the signal amplification, we show the measured position displacement $\Delta Q$ of the phase-space distribution as a function of $\tau$ in Fig.~\ref{fig:eval}(a) for three different values of $r$. 
For direct comparison, for the conventional protocol ($r=1$), we convert the position displacement into the corresponding momentum displacement reached after a quarter oscillation period.
We find that, for all values of $r$, the scaling of $\Delta Q$ with $\tau$ is linear, as expected from the linear scaling of the transferred momentum $\Delta P$ with $\tau$.
Furthermore, we observe that the vertical offset of the data in Fig.~\ref{fig:eval}(a) grows with the squeezing ratio $r$. Due to the double-logarithmic representation, this offset represents the linear amplification factor $k(r)$ that translates the impulsive force (quantified by $\tau$) into a final position displacement $\Delta Q=k(r)\tau$. 
We plot the extracted values of the amplification factor $k(r)$ against the applied squeezing factor $r$ in Fig.~\ref{fig:eval}(b) and observe a linear scaling according to $k(r)=k_1\, r$.
The proportionality constant $k_1$ is determined by the electrode geometry, the applied voltage, and the particle's charge~\cite{supplement}. 

\paragraph{Noise analysis.}
We now turn to the analysis of the noise in our protocol. Experimentally, it is the variance of the marginal distribution along the position axis $\sigma_{Q,\text{tot}}^2$ after the anti-squeezing step, as illustrated in Figs.~\ref{fig:timetraces}(f,\,h).
In Fig.~\ref{fig:eval}(c), we plot the measured values of $\sigma_{Q,\text{tot}}$ as a function of $r$. We observe that $\sigma_{Q,\text{tot}}$ has a constant offset of about 2.5, and grows slightly with $r$.

According to our model, the variance $\sigma_{Q,\text{tot}}^2=\sigma_{Q,i}^2+\sigma_{Q,f}^2$ has two contributions [c.f. Fig.~\ref{fig:concept}(b)]. 
The first contribution $\sigma_{Q,i}=\sigma_{P,i}$ stems from the initialization of the oscillator before the impulsive-force arrives. It can be expressed using the phonon population $n$ of the feedback-cooled oscillator as $\sigma_{P,i}=\sqrt{2n+1}$. 
In our case ($n=1.2$), we expect $\sigma_{Q,i}^2=3.4$.
The second contribution is given by the noise associated with the estimation of the final state $\sigma_{Q,f}^2$. In our case of an efficient quantum measurement, this noise is given by the detection efficiency $\eta$ according to $\sigma^2_{Q,f}=\eta^{-1/2}=2.7$ for our system. 
The expected total uncertainty $\sigma_{Q,\text{tot}}=2.5$ is in good agreement with our observation at small values of $r$ in Fig.~\ref{fig:eval}(c). 
The additionally observed linear growth of $\sigma_{Q,\text{tot}}$ with $r$ is quantitatively explained by photon recoil heating during the squeezing and anti-squeezing steps, whose duration increases with increasing $r$ (see Supplemental Material~\cite{supplement}).

\paragraph{Sensitivity analysis.}
\begin{figure}[b]
\hspace{-2ex}
\includegraphics[width=\columnwidth]{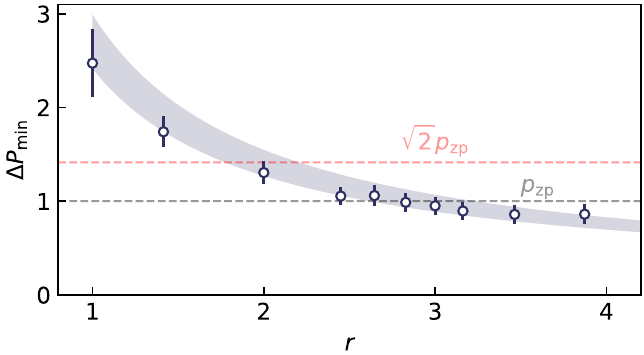}
\caption{\label{fig:pmin}
Minimum detectable impulsive force $\Delta P_{\text{min}}$ as a function of  squeezing ratio $r$. 
Red dashed line: Limit of the conventional protocol for an otherwise ideal system. The label $\sqrt{2}p_\text{zp}$ is expressed in absolute units. 
Grey dashed line: Value of the momentum zero-point fluctuations of the levitated oscillator (label in absolute units). 
Error bars: $1\sigma$ uncertainty obtained from error propagation.
Shaded area: $1\sigma$ confidence interval of the model~\cite{supplement}.
}
\end{figure}
How small of an impulsive force can our system resolve?
To be detectable in a single iteration of the experiment, the impulsive force $\Delta P_\text{min}$, when amplified by our protocol and transferred to the position quadrature, must at least equal the total position uncertainty of the final oscillator state, yielding $\Delta P_\text{min}=\sigma_{Q,\text{tot}}/r$.
In Fig.~\ref{fig:pmin}, we use the data from Figs.~\ref{fig:eval}(b,c) to plot $\Delta P_\text{min}$. The observed suppression of $\Delta P_\text{min}$ for increasing $r$ is the result of the signal amplification by the reversible squeezing operation in the presence of minimally added noise. The smallest value of $\Delta P_\text{min}$ reached, in absolute units, is $6.9\pm0.8$~keV/c.

As a quantitative benchmark, let us compare our system performance to the value of $\Delta P_\text{min}$ achievable without the use of coherent amplification, but with an otherwise optimal system (no excess backaction and detection efficiency $\eta=1$). In that case, we have $\sigma_{P,i}=\sigma_{P,f}=1$, such that the minimum detectable impulsive force is $\Delta P_\text{min}=\sqrt{2}$ [red dashed line in Fig.~\ref{fig:pmin}]. This result matches the quantum limit of continuous force sensing~\cite{Clerk_PRB_2004_quantumLimited}. 
Thus, using coherent mechanical amplification, our system provides a $\Delta P_\text{min}$ that is 2.1~dB smaller than that of the conventional protocol executed with an ideal system. 
Remarkably, the sensitivity reached by our experiments is even $0.6^{+0.6}_{-0.4}$~dB smaller than a single zero-point fluctuation $p_\text{zp}$ of our oscillator, shown as the dashed gray line in Fig.~\ref{fig:pmin}. 

Our current sensitivity is primarily constrained by the squeezing ratio $r$. Excessively large values of $r$ shift the particle's oscillation frequency into a frequency band with significant mechanical vibrations, leading to increased excess backaction. 
Furthermore, the weak confinement during the (anti-)squeezing sequence associated with large $r$ poses the risk of particle loss.

\paragraph{Conclusion.}
We have demonstrated coherent mechanical amplification of impulsive forces acting on an optically levitated nanoparticle. 
The signal amplification has been achieved by embedding the arrival of the impulsive force within a squeezing--anti-squeezing sequence which temporarily redistributes the zero-point fluctuations from the oscillator's momentum to its position.
We have demonstrated an impulsive-force sensitivity that is 2.1~dB smaller than the quantum limit achievable without reversible squeezing using an otherwise ideal system.

A potential way to push the sensitivity of our system further could be to use hybrid traps, where the low-stiffness potential required for (anti-)squeezing is provided by a radio-frequency field and is extremely deep~\cite{Bonvin_PRL_2024_hybridtrap,Conagla_nanoletters_2020_hybridtraps,Bykov_2022}. 
Such dark potentials would also remedy the residual quantum backaction in optical traps. 
Furthermore, powerful methods other than frequency jumps have recently been developed to squeeze the state of a levitated nanoparticle, including free-falls~\cite{mattana_arxiv_2025_freefalls, Steiner_2025_APL_chargedfreefalls,Kamba2022optical,hebestreit_PRL_2018_freefals} and evolution in an inverted potential~\cite{Tomassi_PRR_2025_acceleratedexpansion, Seta2025_shot2shotInverted}. Once amended with an anti-squeezing step, these methods could be attractive candidates to further boost the attainable amplification.

The technique demonstrated in this work could contribute to scientific discoveries on different fronts. 
Our method may enable detection of unknown elementary particles~\cite{moore_QST_2021_newphysics}, such as various dark matter candidates \cite{Moore_PRL_2020_compDM,Moore_2025_PRXQ_DMsearch}, neutrinos~\cite{Carney_PRXQ_2023_neutrinosearch}, or particles created in nuclear decays~\cite{wang_PRL_2024_nucleardecays}. In a more applied direction, quantum-enhanced sensing with levitated particles may benefit the analysis of rarefied media by collision measurements~\cite{Gajewski2025_LEVITAS,badurina_PRA_2024_collisionaldecoherence}.

\subsection*{Acknowledgments}

We thank Oscar Schmitt Kremer for his help with the Kalman filter and the rest of our colleagues at the ETH Photonics Laboratory for fruitful discussions. This research has been supported by the Swiss SERI Quantum Initiative (grant no. UeM019-2), the Swiss National Science Foundation (grant no. 51NF40-160591), and the European Research Council (ERC) under the grant agreement No. [951234] (Q-Xtreme ERC-2020-SyG). M.C.S. thanks for support through an SNSF Fellowship (grant no. 224465). \\

\bibliography{skrabulis_bib}%

\end{document}